\documentclass[12pt]{article}

\usepackage{graphicx,psfrag,epsf}
\usepackage{enumerate}
\usepackage{natbib}
\usepackage{mathtools}
\usepackage{booktabs}
\usepackage{color}
\usepackage[english]{babel} 
\usepackage[protrusion=true,expansion=true]{microtype} 
\usepackage{amsmath,amsfonts,amsthm}
\usepackage{dsfont}
\usepackage{amssymb}
\usepackage{chngcntr}
\usepackage[toc,page]{appendix}
\usepackage{textcomp}
\usepackage{url}

\usepackage{array}
\usepackage{booktabs} 
\usepackage{natbib}
\usepackage{setspace}

\usepackage{soul}
\usepackage{multirow}
\usepackage{enumitem}
\usepackage{microtype} 
\usepackage[font = small,labelfont=bf,textfont=it]{subcaption}
\usepackage{xcolor}
\usepackage{tabularx}
\usepackage[font = small,labelfont=bf,textfont=it]{caption} 
\usepackage{footnote}
\usepackage{algorithm}
\usepackage[noend]{algpseudocode}
\usepackage{etoolbox}

\algblock{ParFor}{EndParFor}
\algnewcommand\algorithmicparfor{\textbf{for}}
\algnewcommand\algorithmicpardo{\textbf{do\ parallel}}
\algnewcommand\algorithmicendparfor{\textbf{end\ parallel\ for}}
\algrenewtext{ParFor}[1]{\algorithmicparfor\ #1\ \algorithmicpardo}
\algrenewtext{EndParFor}{\algorithmicendparfor}

\makeatletter
\def\BState{\State\hskip-\ALG@thistlm}

\newcommand{\distas}[1]{\mathbin{\overset{#1}{\kern\z@\sim}}}%
\newcommand{\bm}[1]{\mathbf{#1}}
\newcommand{\bb}[1]{\boldsymbol{#1}}
\newsavebox{\mybox}\newsavebox{\mysim}
\newcommand{\distras}[1]{%
  \savebox{\mybox}{\hbox{\kern3pt$\scriptstyle#1$\kern3pt}}%
  \savebox{\mysim}{\hbox{$\sim$}}%
  \mathbin{\overset{#1}{\kern\z@\resizebox{\wd\mybox}{\ht\mysim}{$\sim$}}}%
}

\newcommand{\be}{\begin{equation}}
\newcommand{\ee}{\end{equation}}
\newcommand{\bi}{\begin{itemize}}
\newcommand{\ei}{\end{itemize}}
\newcommand{\ben}{\begin{enumerate}}
\newcommand{\een}{\end{enumerate}}

\newcolumntype{K}[1]{\geq {\centering\arraybackslash}p{#1}}
\allowdisplaybreaks
\DeclareMathOperator*{\argmin}{\arg\!\min}
\makeatother

\let\oldbibliography\thebibliography
\renewcommand{\thebibliography}[1]{\oldbibliography{#1}
\setlength{\itemsep}{0pt}} 

\newcommand{\blind}{0}

\patchcmd{\footnotemark}{\stepcounter{footnote}}{\refstepcounter{footnote}}{}{}

\usepackage{booktabs}

\usepackage{makecell}
\usepackage[section]{placeins}

\begin{document}

\def\spacingset#1{\renewcommand{\baselinestretch}%
{#1}\small\normalsize} \spacingset{1}

\if1\blind
{
  \title{\bf }
  \small
  \author{}\hspace{.2cm}\\
  \maketitle
} \fi

\if0\blind
{
  \bigskip
  \bigskip
  \bigskip
  \begin{center}
    {\LARGE\bf  Discovering the Unknowns: A First Step}\vspace{.2cm}\\
    {V. Roshan Joseph$^1$, William E. Lewis$^2$, Henry S. Yuchi$^1$, and Kathryn A. Maupin$^2$}\vspace{.2cm}\\
    $^1${H. Milton Stewart School of Industrial and Systems Engineering\\ 
    Georgia Institute of Technology, Atlanta, GA 30332, USA}\vspace{.2cm}\\
    $^2${Sandia National Laboratories, Albuquerque, NM 87185, USA}
\end{center}
  \medskip
} \fi
\bigskip

\vspace{-0.5cm}
\begin{abstract}
\noindent  This article aims at discovering the unknown variables in the system through data analysis. The main idea is to use the time of data collection as a surrogate variable and try to identify the unknown variables by modeling gradual and sudden changes in the data. We use Gaussian process modeling and a sparse representation of the sudden changes to efficiently estimate the large number of parameters in the proposed statistical model. The method is tested on a realistic dataset generated using a one-dimensional implementation of a Magnetized Liner Inertial Fusion (MagLIF) simulation model and encouraging results are obtained.
\end{abstract}


\spacingset{1.5} 

\section{Introduction} \label{sec:intro}
In an attempt to develop a relationship between the output ($y$) and inputs ($\bm x$) from data $\{(\bm x_i,y_i)\}_{i=1}^n$, it is common to start by postulating a statistical model of the form
\begin{equation}\label{eq:model0}
    y_i=f(\bm x_i)+\epsilon_i,
\end{equation}
where $\epsilon_i$ is the random noise in the output. The random noise is caused by the variability of the ``unknown'' variables in the system and the measurement error. The unknown variables (also known as hidden or lurking variables) could be the variables ignored during the data collection thinking that they are unimportant or the variables that are truly unknown to the investigator. The random noise is usually modeled using a stochastic process, for example, it is common to assume $\epsilon_i\overset{iid}{\sim} N(0,\sigma^2)$. In this modeling framework, one focuses on estimating the functional relationship ($f$) between $y$ and the ``known'' set of variables $\bm x=(x_1,\ldots,x_p)'$ using parametric or nonparametric regression methods and the parameters of the noise distribution ($\sigma^2$). Different from this common approach, the aim of this article is to understand or discover the unknown variables contributing to the random noise term.

The random error term is thought to be caused by numerous variables in the system that are difficult to parse and thus is ignored as a ``noise'' in the modeling framework. They are considered to be caused by natural variability in the system and therefore, are put under the umbrella of aleatoric uncertainty. Nothing much is done on the noise term other than trying to characterize its probability distribution. An exception is the robust parameter design approach \citep{taguchi1987system}, where the noise variability $\sigma^2$ is also modeled as a function of the known variables $\bm x$ in order to find a setting of $\bm x$ that will reduce the effect of noise. Different from Taguchi's approach and other heteroscedastic statistical modeling methods \citep{mccullagh1989generalized}, our aim is to understand the causes of the noise. If we can successfully identify the causes (unknown variables), they can be added to the model and thus, reduce the modeling error in (\ref{eq:model0}).

In observational studies, the presence of the unknown variables can create doubts about the causality of the input variables under study. Therefore, the causal inference techniques \citep{imbens2015causal} focus on verifying the causality of the known input variables. On the other hand, our focus here is on discovering the unknown variables. If we could not identify any important unknown variables, then it might help in indirectly establishing the causality of the known variables.


Establishing the relationship of the output with the known variables itself is a difficult task, so it is clear that dealing with the ``unknowns'' is going to be an even more challenging task. In this article we will propose an approach to discover the unknown variables using Gaussian Process (GP) modeling \citep{santner2003design, rasmussen2006gaussian}. It should be viewed only as a first step towards the discovery process. Follow-up investigations are necessary to truly identify the variables, but that will not be addressed in this article.

The article is organized as follows. Section 2 develops the statistical methodology for identifying the unknown variables. The proposed method is tested using a toy example in Section 3 and on a realistic dataset generated using a one-dimensional implementation of a Magnetized Liner Inertial Fusion (MagLIF) simulation model in Section 4. Some concluding remarks and future research directions are given in Section 5.

\section{Methodology}
The ultimate aim of our methodology is to develop a model of the form
\begin{equation}\label{eq:model1}
    y=f(\bm x, \bm u)+\epsilon',
\end{equation}
where $\bm u=(u_1,\ldots,u_q)$ are the newly discovered variables which were unknown before the study and $\epsilon'$ is the random noise, which is now only due to the measurement error. This may be impossible to achieve as there could be numerous $u_i$'s. Moreover, some of the $u_i$'s may have negligible effects on the output and therefore, it may not be worth including them in the model. Thus, we will be satisfied if we can at least identify a few of the $u_i$'s that have major impact on the output.

Our main idea is to classify the $u_i$'s into two categories: \emph{gradual} and \emph{sudden}. As the names imply, the $u_i$'s in the first category will cause gradual changes in the output, whereas the $u_i$'s in the second category will cause sudden changes in the output. For example, some variables may be changing over time, the system or its components may be degrading over time, there could be wear and tear happening to some parts, there could be changes to the environment (temperature and humidity), there could be accumulation of impurities over time, and so on. These are time-dependent changes, which we classify as ``gradual''. The second type includes change of material batches, change of machines, change of operators, setup changes, cleaning, etc. These cause immediate changes in the output. 

We only have the data $(\bm x_1,y_1),\ldots,(\bm x_n,y_n)$ and no additional information on the response surface or the unknown variables. Thus, using the data we need to find out if there are any gradual  or  sudden effects. Once we identify this, brainstorming or follow-up experiments may help to pinpoint the real missing variables.

Assume that we have recorded the time ($t$) of system operation. Then, we will attempt to discover the gradual effects using ``time'' as a surrogate variable and the sudden effects using random effects $\delta_i$. Thus, we propose to model the output as
\begin{equation}\label{eq:propmodel}
    y_i=f(\bm x_i, t_i)+\delta_i+\epsilon_i,\;\;i=1,\ldots,n,
\end{equation}
where $\epsilon_i\overset{iid}{\sim} N(0,\sigma^2)$. The random effects $\delta_i$ are suppose to correspond to a setup, batch, machine, etc., but this is not known beforehand. Thus, we introduce a $\delta_i$ for every observation. This formulation has some similarities to the smooth-sparse anomaly detection techniques used in the process control literature \citep{yan2017anomaly}, but their problem setting is quite different from ours.

Given the data $\{(\bm x_i,y_i)\}_{i=1}^n$, our aim is to estimate $f$ and $\{\delta_i\}_{i=1}^n$. We will use a Bayesian approach by postulating priors for these quantities. Gaussian processes (GPs) are widely used as priors for functions \citep{santner2003design}. So we will let $f$ to be a realization from a GP. To simplify the notations, let $x_{p+1}=t$ and $\tilde{\bm x}=(\bm x,t)$. Thus, assume
\begin{equation}
    f(\tilde{\bm x})\sim GP(\mu,\tau^2R(\cdot;\bb \theta)),
\end{equation}
where $\mu$ is the mean, $\tau^2$ the variance,  $R(\tilde{\bm x}_i-\tilde{\bm x}_j;\bb \theta)=cor\{f(\tilde{\bm x}_i),f(\tilde{\bm x}_j)\}$ the stationary correlation function, and $\bb \theta$ the correlation parameters. 

\begin{figure}[h]
\begin{center}
\includegraphics[width = .65\textwidth]{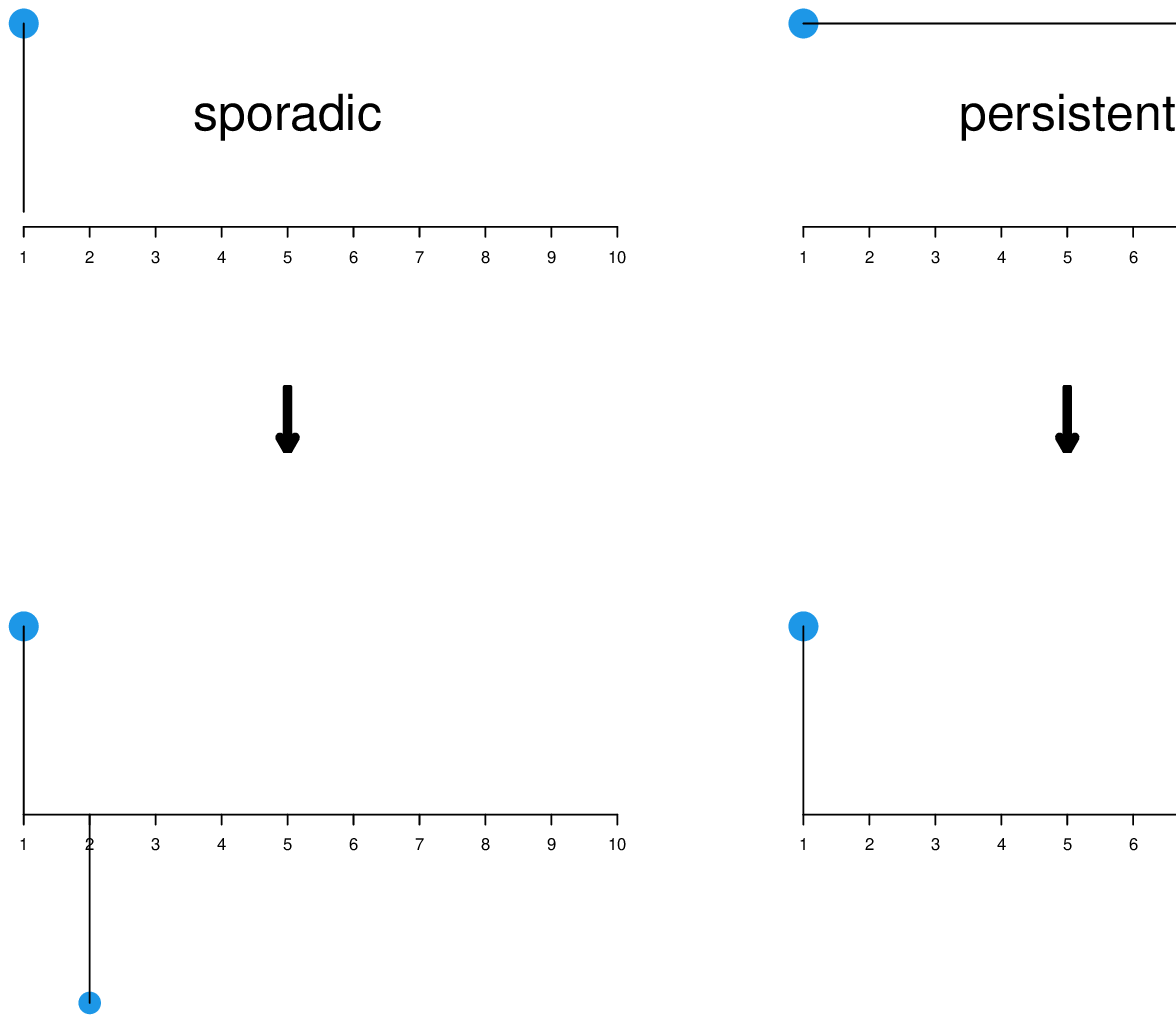} 
\caption{Representation of sporadic and persistent changes (top row) using $e_i$'s with the $\bm U$ matrix as in (\ref{eq:U}) (bottom row).}
\label{fig:sudden}
\end{center}
\end{figure}

Now consider the prior specification on the $\{\delta_i\}_{i=1}^n$. There are too many parameters to estimate (as large as the number of data points). To simplify this estimation problem, we will develop a sparse representation of $\bb \delta=(\delta_1,\ldots,\delta_n)'$ as follows:
\begin{equation}
    \bb \delta=\bm U \bm e,
\end{equation}
where $\bm e=(e_1,\ldots,e_n)'$. Now we will choose $\bm U$ in such a way that only a few $e_i$'s will be non-zero. In general, two types of sudden changes can occur: \emph{sporadic} or \emph{persistent}. The effect of sporadic changes are instantaneous and will not last long. For example, the changes in material batches or machines will not have a carry over effect. On the other hand, a system tune-up or cleaning can bring the system to almost as good as new and its effect stays in the system. We will call the former type of effects as sporadic and the latter type as persistent. If we know a priori that the effects are sporadic, then choosing $\bm U=\bm I$, the identity matrix, is the best. However, if we know the effects to be persistent, then we should choose 
\begin{equation}\label{eq:U}
    \bm U=\begin{bmatrix}
    1 &0 &0 &\ldots &0\\
    1 &1 &0 &\ldots &0\\
    \vdots &\vdots &\vdots &\vdots &\vdots\\
    1 & 1 &1 &\vdots &1
    \end{bmatrix}.
\end{equation}
Unfortunately, we do not know if the sudden effects are going to be sporadic or persistent. A sporadic change can be represented using two $e_i$'s if we choose the $\bm U$ in (\ref{eq:U}), whereas a persistent change that lasts for 10 time steps would require 10 $e_i$'s if we were to use $\bm U=\bm I$. See Figure \ref{fig:sudden} for an illustration. Since we do not know if the effects are going to be sporadic or persistent a priori, it is best to use the $\bm U$ in (\ref{eq:U}). This choice is likely to provide us with a sparse representation of the sudden changes.

Now we can postulate a prior on $e_i$ so that it will be 0 most of the time and becomes active when a sudden change happens in the system. Thus, we assume that $e_i$ follows a double exponential  distribution (also known as Laplace distribution):
\begin{equation}
    e_i\sim^{iid} DE(0,\nu),
\end{equation}
where the density of $e$ is given by $p(e)=1/(\sqrt{2}\nu)\exp\{-\sqrt{2}|e|/\nu\}$ and $\nu$ denotes the standard deviation. We will later see that this prior distribution on $e_i$ leads to a sparse solution. \cite{chang2014model} used a similar idea in model calibration, but their aim was to adjust the physics-based model based on data by making minimal changes to it, which is completely different from our objective.

Thus, given the data $\{(\bm x_i,y_i)\}_{i=1}^n$, our aim is to find the posterior distribution of $f(\tilde{\bm x})$ and the $e_i$'s for $i=1,\ldots,n$. For this, we need to estimate all the unknown quantities: $\bb \phi=(\mu,\tau^2,\nu,\sigma^2,\bb \theta)$ from the data (or integrate them out). Since the first column of $\bm U$ is a column of $1$'s, $e_1$ is unidentifiable from $\mu$. Therefore, we will let $e_1=0$.

Let $\bm y=(y_1,\ldots,y_n)'$, $\bm e=(\mu,e_2,e_3,\ldots,e_n)'$, $\bm R$ the correlation matrix whose $ij$th element is $R(\bm x_i-\bm x_j;\bb \theta)$, and $\bm I$ the $n\times n$ identity matrix. Since $\bm y|\bb \phi, \bm e \sim N(\bm U\bm e, \tau^2\bm R+\sigma^2\bm I)$, the unnormalized posterior of $\bb \phi$ and $\bm e$ is given by
\begin{eqnarray}
    p(\bb \phi,\bm e|\bm y)&\propto & \frac{1}{|\tau^2\bm R+\sigma^2\bm I|^{1/2}}\exp\left\{-\frac{1}{2}(\bm y-\bm U\bm e)'(\tau^2\bm R+\sigma^2\bm I)^{-1}(\bm y-\bm U\bm e)\right\}\nonumber\\
    &&\times\frac{1}{\nu^n}\exp\left\{-\frac{\sqrt{2}}{\nu}\sum_{i=2}^n|e_i|\right\},\label{eq:EB}
\end{eqnarray}
It is not possible to analytically integrate out $\bm e$ from this expression. However, given $\bb \phi$, we can estimate $\bm e$ by maximizing
\[-\frac{1}{2\tau^2}(\bm y-\bm U\bm e)'(\bm R+\eta\bm I)^{-1}(\bm y-\bm U\bm e)-\frac{\sqrt{2}}{\nu}\sum_{i=2}^n|e_i|,\]
where $\eta=\sigma^2/\tau^2$ is the ``nugget''. Equivalently, $\bm e$ can be obtained by minimizing
\begin{equation}\label{eq:lasso}
    (\bm y-\bm U\bm e)'(\bm R+\eta\bm I)^{-1}(\bm y-\bm U\bm e)+\lambda\sum_{i=2}^n|e_i|,
\end{equation}
where $\lambda=2\sqrt{2}\tau^2/\nu$, which can be solved efficiently using the well-known algorithms for lasso \citep{tibshirani1996regression}, provided $\eta$ and $\bb \theta$ are known. 

We can estimate $\lambda$ using cross validation, which will give the desired estimates of $\bm e$. A fully Bayesian analysis could be employed to get the posterior distribution of $\lambda$ and $\bm e$ \citep{park2008bayesian}, but we contend here with their point estimates for the sake of computational efficiency. Denote them as $\hat{\lambda}$ and $\hat{\bm e}$, respectively. By plugging them into (\ref{eq:EB}) and maximizing, we obtain
\begin{eqnarray}
    \hat{\tau}^2&=&\frac{1}{3n}\left\{(\bm y-\bm U\hat{\bm e})'(\bm R+\eta\bm I)^{-1}(\bm y-\bm U\hat{\bm e})+\hat{\lambda}\sum_{i=2}^n|\hat{e}_i|\right\},\label{eq:tau2}\\
    \hat{\bb \theta},\hat{\eta}&=&\argmin_{\bm \theta,\eta}\; 3n\log \hat{\tau}^2+\log |\bm R+\eta\bm I|.\label{eq:theta}
\end{eqnarray}
Equations (\ref{eq:lasso}),(\ref{eq:tau2}), and (\ref{eq:theta}) need to be solved iteratively, which makes the optimization in (\ref{eq:theta}) computationally demanding especially when $n$ is large. Another issue with the optimization is that the objective function is stochastic due to the cross validation-based selection of $\lambda$ in the lasso algorithm. We can remove this randomness by fixing the folds for cross validation outside the optimization loop. Furthermore, a good choice of the folds can be obtained using the twinning algorithm of \cite{vakayil2022data}.


The posterior distribution of $f(\tilde{\bm x})$ is given by
\begin{equation}
    f(\tilde{\bm x)})|\bm y, \hat{\bm e}\sim N\left(\hat{f}(\tilde{\bm x}), s^2(\tilde{\bm x})\right),
\end{equation}
where 
\begin{eqnarray}
    \hat{f}(\tilde{\bm x})&=&\hat{\mu}+\bm r(\tilde{\bm x})'(\bm R+\hat{\eta} \bm I)^{-1}(\bm y-\bm U\hat{\bm e}),\label{eq:fhat}\\
    s^2(\tilde{\bm x})&=&\tau^2\left\{1- \bm r(\tilde{\bm x})'(\bm R+\hat{\eta} \bm I)^{-1} \bm r(\tilde{\bm x})\right\}\nonumber.
\end{eqnarray}
Integrating out $\bm x$ from (\ref{eq:fhat}), we obtain the gradual effect as
\[g(t)=\hat{\mu}+\int \bm r(\tilde{\bm x})'\;d\bm x\;(\bm R+\hat{\eta} \bm I)^{-1}(\bm y-\bm U\hat{\bm e}).\]
Assume a product correlation structure between $\bm x$ and $t$, that is, $R(\tilde{\bm x}-\tilde{\bm x}_i;\bm \theta)=R_x(\bm x-\bm x_i;\bm \theta_x)R_t(t-t_i;\theta_t)$.
Then,
\begin{equation}\label{eq:gradual}
    g(t)=\hat{\mu}+(\bm r(t)\odot \bm d_x)'(\bm R+\hat{\eta} \bm I)^{-1}(\bm y-\bm U\hat{\bm e}),
\end{equation}
where $(\bm d_x)_i=\int R_x(\bm x-\bm x_i;\bm \theta_x)\;d\bm x$. Explicit expressions for $\bm d_x$ can be obtained for some specific correlation functions. For example, if $R_x(\bm x-\bm x_i;\bm \theta_x)=\exp\{-\sum_{k=1}^p \theta_{x,k} (x_k-x_{ik})^2\}$, then (assuming $\bm x$ is scaled in $[0,1]^p$)
\[(\bm d_x)_i= \prod_{k=1}^p \sqrt{\frac{\pi}{\theta_{x,k}}}\left\{\Phi\left(\sqrt{2\theta_{x,k}}(1-x_{ik})\right)-\Phi\left(-\sqrt{2\theta_{x,k}}x_{ik}\right)\right\},\]
where $\Phi(\cdot)$ is the standard normal distribution function.

Once we obtain $g(t)$, we can plot $g(t)$ as well as $\bm U\hat{\bm e}$ over time and look for patterns. If some patterns exists, they might give clues regarding the unknown variables in the system. We will illustrate this idea with some examples in the next two sections.

\section{Simulations}
\begin{figure}[h]
\begin{center}
\includegraphics[width = .6\textwidth]{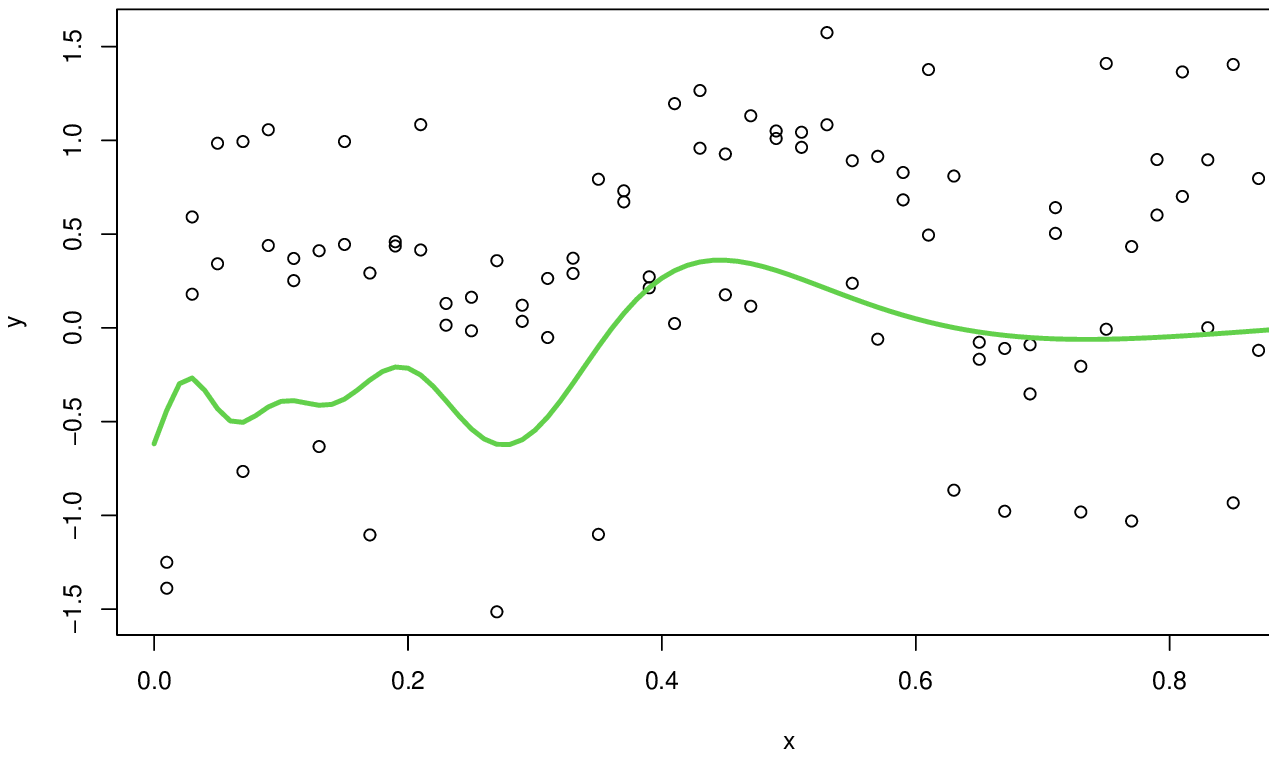} 
\caption{Data and the true function (green line) in the example.}
\label{fig:data_example}
\end{center}
\end{figure}

Consider a one-dimensional test function used in \cite{xiong2007non}:
\[ f(x)=sin(30(x-0.9)^4)cos(2(x-0.9))+(x-0.9)/2, \;x\in [0,1].\]
Let $x_i=(i-.5)/50$, $i=1,2,\ldots,50$, with each point replicated twice. The 100 values are completely randomized and the data are generated by adding a linear trend over time and five sudden (persistent) changes. Specifically,
\[y_i=f(x_i)+2(t_i-0.5)+\sum_{k=1}^i e_i+\epsilon_i,\]
where $t_i=(i-1)/99$, $i=1,2,\ldots,100$ and $e_i=\pm 0.5$ for $i=10,30,50,70,90$ (sign chosen randomly) and 0 elsewhere, and $\epsilon_i\overset{iid}{\sim} N(0,.01^2)$. The data are plotted in Figure \ref{fig:data_example} with the true function overlaid.

\begin{figure}[h!]
\begin{center}
\includegraphics[width = .99\textwidth]{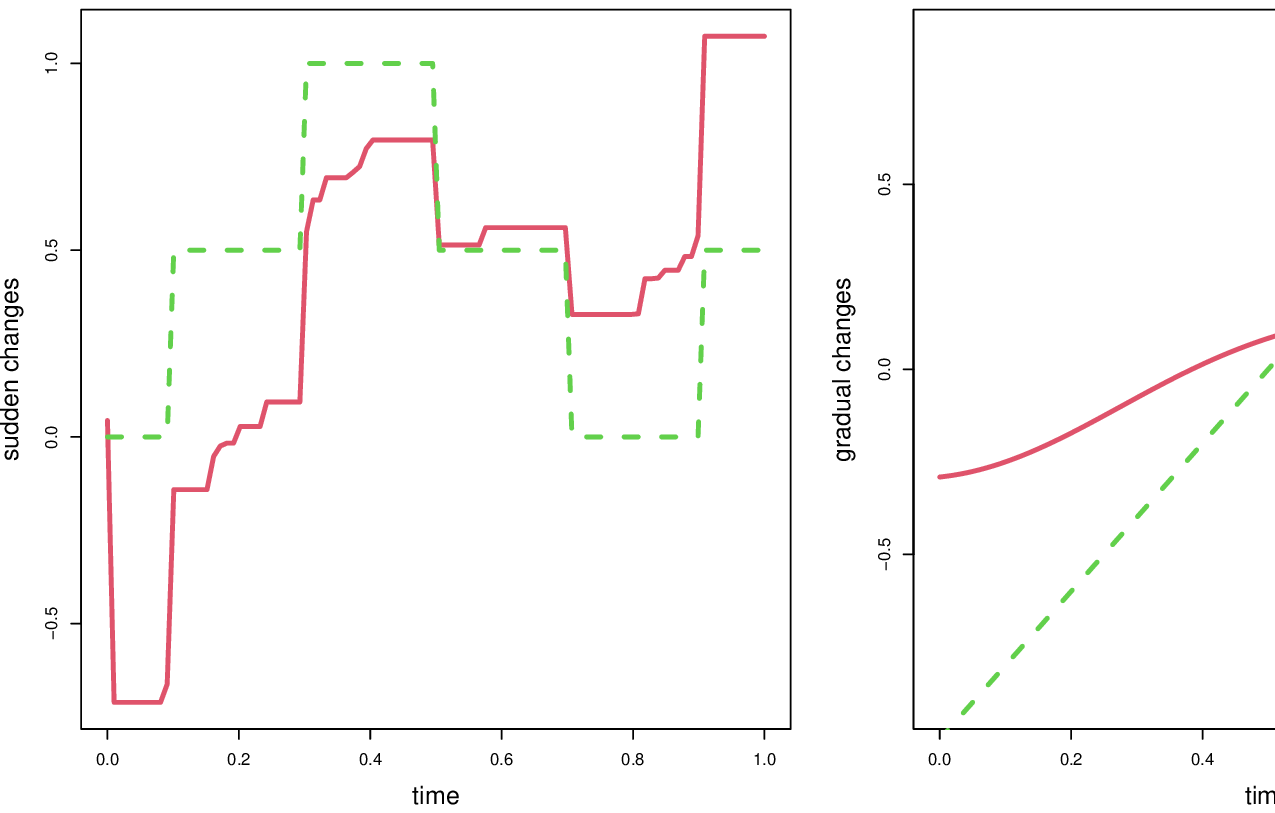} 
\caption{Sudden changes (left), gradual changes (middle), and input-output relationship (right). The true lines are shown as green dashed lines.}
\label{fig:est_example}
\end{center}
\end{figure}

Following the procedure developed in Section 2, we first estimate $\theta$, $\eta$, $\tau^2$, and $\{e_i\}_{i=2}^n$ using (\ref{eq:lasso}),(\ref{eq:tau2}), and (\ref{eq:theta}) assuming a Gaussian correlation function for both $x$ and $t$. The estimated sudden changes $\bm U\hat{\bm e}$ against time are plotted in the left panel of Figure \ref{fig:est_example}. We can see that although the estimation is not perfect, it correctly identifies the locations of the sudden changes. The gradual effect is computed using (\ref{eq:gradual}) and is plotted in the middle panel of the same figure. Although underestimated, it correctly identifies the increasing trend, which may be enough to understand the cause behind it. The final estimate of the $f(x)$ is shown on the right panel (red line), which shows reasonable accuracy compared to the true function (green dashed line).

\begin{figure}[h!]
\begin{center}
\includegraphics[width = .99\textwidth]{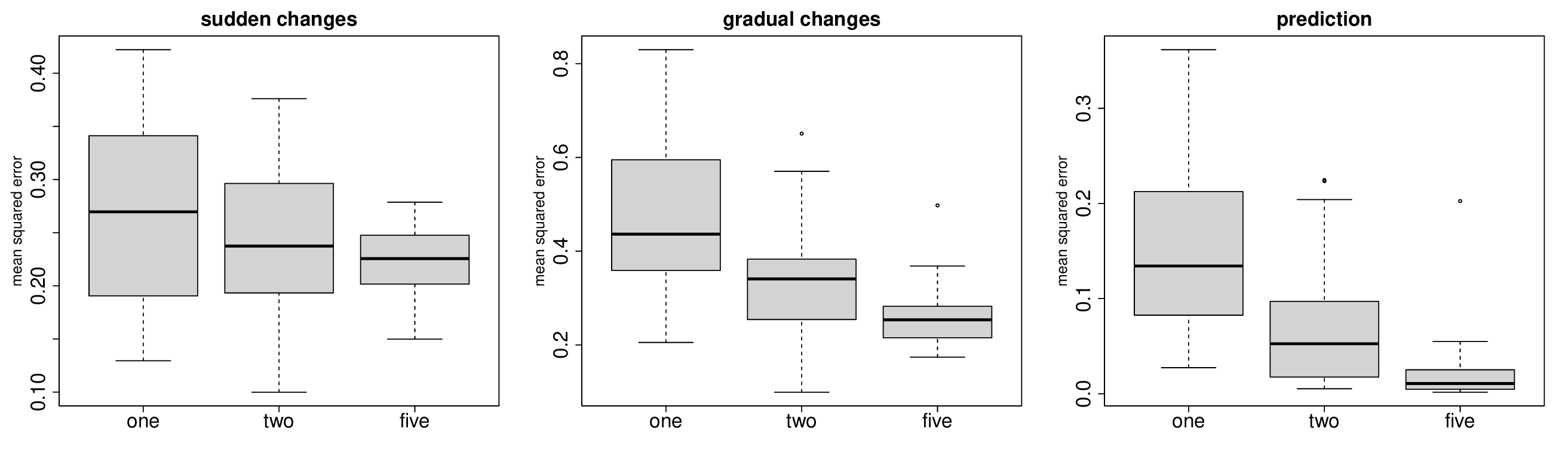} 
\caption{Mean-squared errors of sudden changes (left), gradual changes (middle), and prediction (right) with different number of replications.}
\label{fig:replication}
\end{center}
\end{figure}

The above exercise is repeated 30 times by varying the replications at three levels: $\{1,2,5\}$. The mean squared errors are computed for the sudden, gradual, and prediction profiles (differences between the red and green lines in Figure \ref{fig:est_example}). They are plotted in Figure \ref{fig:replication}. As expected, the estimation improves with increase in the number of replications.

\begin{figure}[h!]
\begin{center}
\includegraphics[width = .99\textwidth]{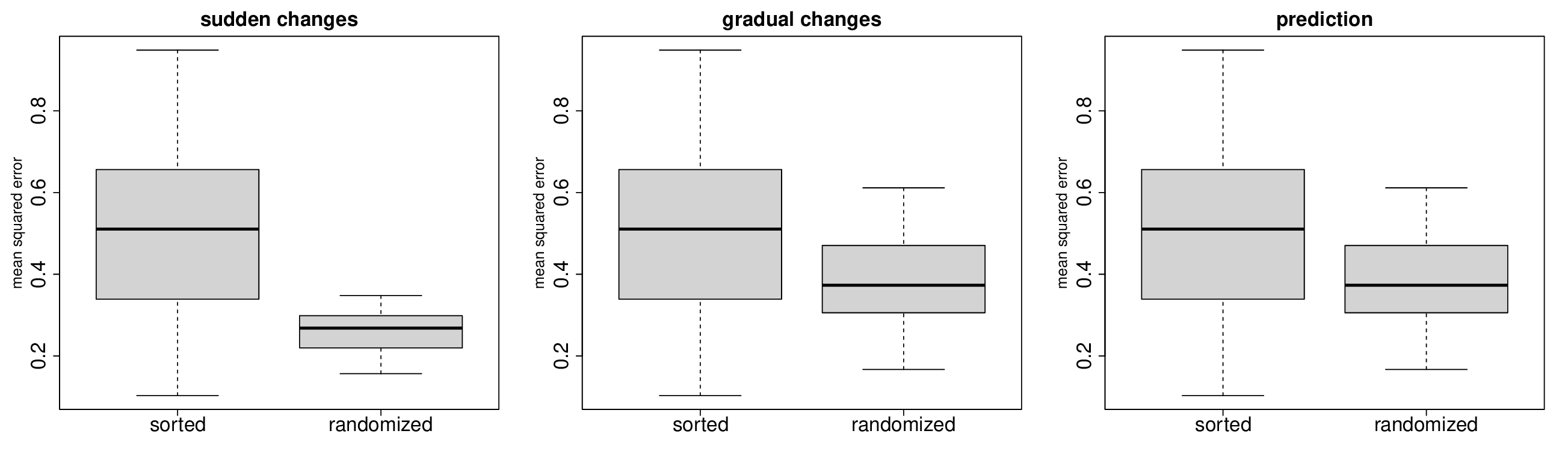} 
\caption{Mean-squared errors of sudden changes (left), gradual changes (middle), and prediction (right) with sorted and randomized $x$ values.}
\label{fig:randomization}
\end{center}
\end{figure}

Now we try another simulation by sorting the $x$ values in increasing order (number of replication is fixed at two). This will make the $x$-values perfectly correlated with the time. Figure \ref{fig:randomization} shows the comparison of mean squared errors for the sorted vs completely randomized case. Again, as expected, the completely randomized case shows improved performance for estimation.

The foregoing two simulations reiterates the importance of replication and randomization in experiments, which are known to be fundamental principles of experimentation \citep{wu-hamada2011}. Clearly, they are crucial for the effectiveness of the proposed procedure for identifying the unknown variables.

\section{Test Problem: Magnetized Liner Inertial Fusion}

\begin{figure}[h]
\begin{center}
\includegraphics[width = .99\textwidth]{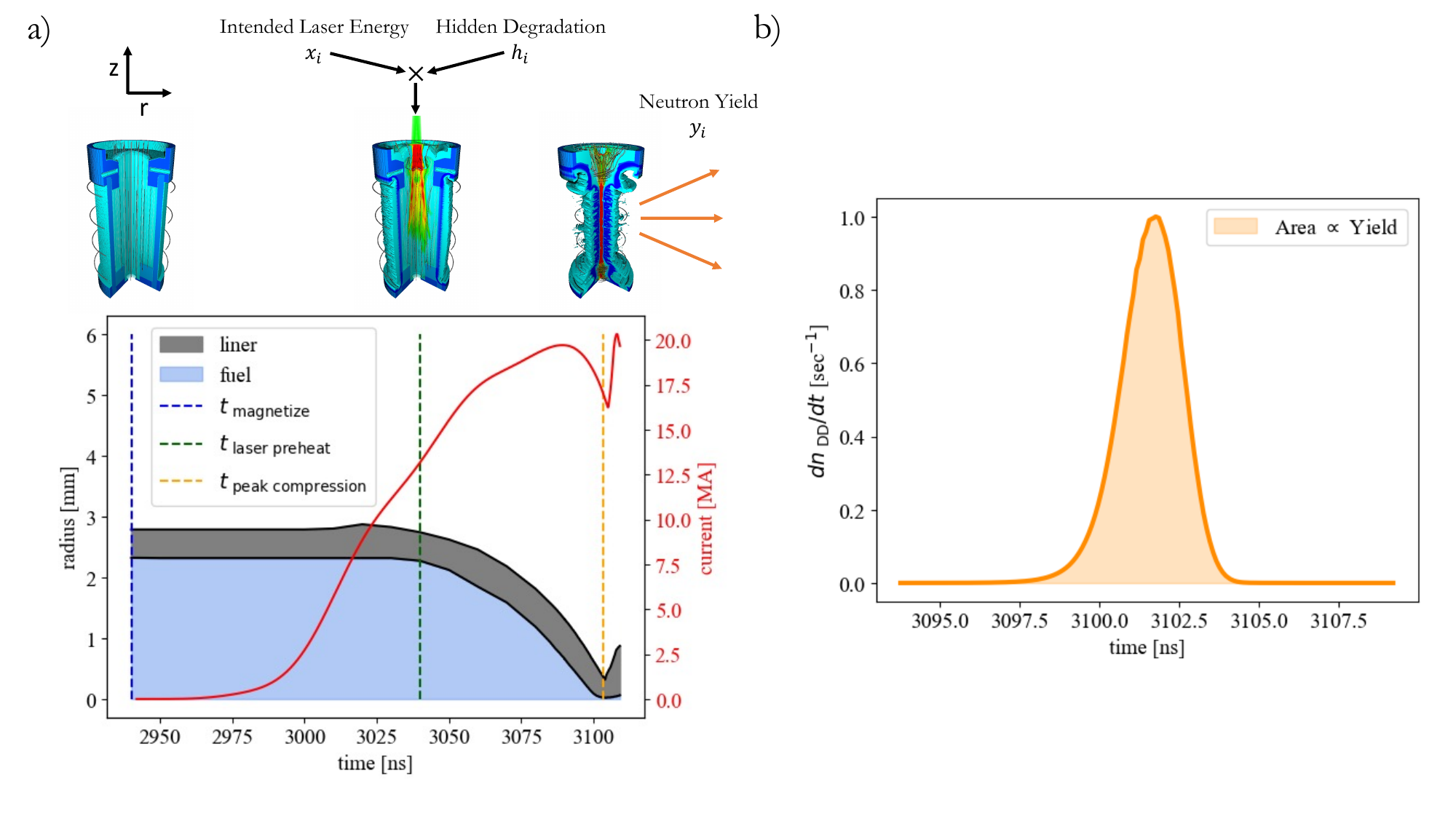} 
\caption{a) The key stages of MagLIF are axial premagnetization, laser preheat, and magnetically driven compression of a cylindrical Beryllium liner containing gaseous deuterium fusion fuel. The included plot shows the trajectory of the inner and outer liner radii as a function of time on the Z-Machine timing basis. The premagnetization serves to thermally insulate the fuel against electron thermal conduction losses throughout the process. The laser serves to increase the fuel temperature early in time so that the compression phase can increase density and temperature to fusion relevant conditions. In the MagLIF exemplar problem, we apply a hidden degradation to the laser energy and attempt to recover information about this unknown variable. }
\label{fig:MagLIF}
\end{center}
\end{figure}

The ability to identify sudden effects from relatively sparse datasets with few replications would prove incredibly useful for costly experiments such as the magnetized liner inertial fusion (MagLIF) concept studied at Sandia's Z facility~\citep{Slutz_PoP_2010,Gomez_PRL_2014,Gomez_PRL_2020}. Briefly, MagLIF experiments proceed through three main steps indicated in Fig.~\ref{fig:MagLIF}. First, an approximately $10$~T axially oriented magnetic field~\citep{Rovang_RSI_2014} is applied to a metallic cylindrical tube or liner filled with pure deuterium fusion fuel to suppress electron thermal conduction losses throughout the experiment. Second, a multi-kilojoule laser~\citep{Rambo_AO_2005,Rambo_SPIE_2016} is injected into the fuel through a top-side laser-entrance window to heat the fuel. This reduces the amount of compression needed to achieve fusion relevant temperatures and densities. Finally, the Z machine~\citep{Savage_IEEE_2008,Rose_PRSTAB_2008,Savage_IEEE_2011} delivers a current pulse that rises to about $20$~MA of peak current in about $100$~ns resulting in a magnetic pressure driving the cylinder to implode.

MagLIF experiments result in a thermonuclear neutron producing plasma that lasts about $2$~ns. This plasma is diagnosed through a wide variety of neutron and x-ray diagnostics, with performance typically measured through the neutron yield (or count). Importantly, it has been observed in some cases that nominally identical experiments can have neutron yields varying by a factor of up to nearly 10. At least in some cases, this variance has been attributed to differences in the growth and feedthrough of the magneto-Rayleigh-Taylor instability  that can disrupt the plasma column and/or introduce liner material into the fuel, enhancing radiative losses, leading to loss of performance~\citep{Awe_PRL_2016,Ampleford_Prep}. In other cases, performance reduction has been hypothesized to result from laser alignment error~\citep{Lewis_PoP_2023}, and may also arise from other off-nominal variations that may go uncharacterized. Motivated by these observations, we create a MagLIF simulation dataset using the 1D resistive magneto-hydrodynamics code Kraken~\citep{Lewis_PoP_2023} to test and validate the methodology presented in this work.  

The test problem data are generated by running an ensemble of 1D Kraken simulations varying only the laser energy deposited. We note here that the laser energy deposition is achieved in the code through a simple heating term that is applied to the gaseous fuel over the area of the laser spot and for the duration of the laser pulse. The dataset includes an additional “measured” piece of information, namely, the number of shots executed without “tune up” that is used to auto-adjust the inputs relative to the specified laser energy. This adjustment is performed using an artificial (multiplicative) degradation factor ($<1$) that depends quadratically on the number of shots executed since the last tune up. The degradation factor is reset to the value 1 upon tune up. The degradation to the energy input into the simulations may be expressed as 
\begin{equation}
    E_{\rm true}(s) = \{1-0.5(s/S)^2\} \times E_{\rm intended},
\end{equation}
where $S=50$ is the number of shots before tune up, $s \in [0,1,2,...,49]$ indicates how many shots have been performed since the last tune up, $E_{\rm intended}$ is the expected energy input, $E_{\rm true}$ is the actual energy provided to the simulation, and tune up consists of resetting the value $s=0$. In this work, the intended energy is varied at $S=50$ levels linearly between $E_{\rm intended} \subset [100~$J,$2500~$J$]$, with two replicates of each energy level giving a total of $n=100$ data points. The replicates are randomly distributed and a tune up is applied after the first 50 simulations.

\begin{figure}[h]
\begin{center}
\includegraphics[width = .6\textwidth]{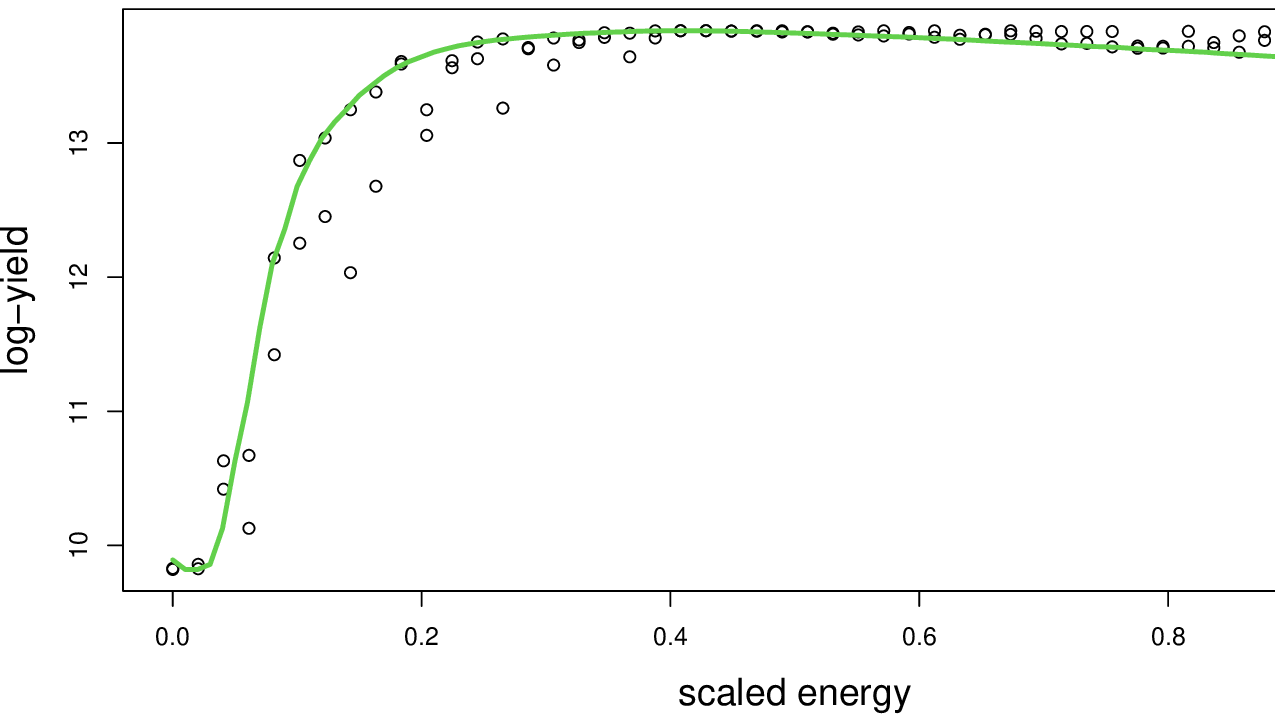} 
\caption{Data and the true function (green line) in the MagLIF simulation.}
\label{fig:data_test}
\end{center}
\end{figure}

Figure \ref{fig:data_test} shows the true relationship between the laser energy (scaled in $[0,1]$) and the yield (in log-scale) as a green line. However, because of the ``degradation'' and ``tune-up'', the actual data are different as shown in  the same figure as black circles. We will analyze this data  without using any knowledge of ``degradation'' and ``tune-up''. As before, the aim is to discover such unknown variables in the system merely through data analysis. We assume a Gaussian correlation function for both the laser energy ($x$) and time ($t$). Following the method presented in Section 2, we obtain the estimates of $\bm \phi=(\mu,\tau^2,\nu,\sigma^2,\bm \theta)$ and $\bm e$.

\begin{figure}[h]
\begin{center}
\includegraphics[width = .99\textwidth]{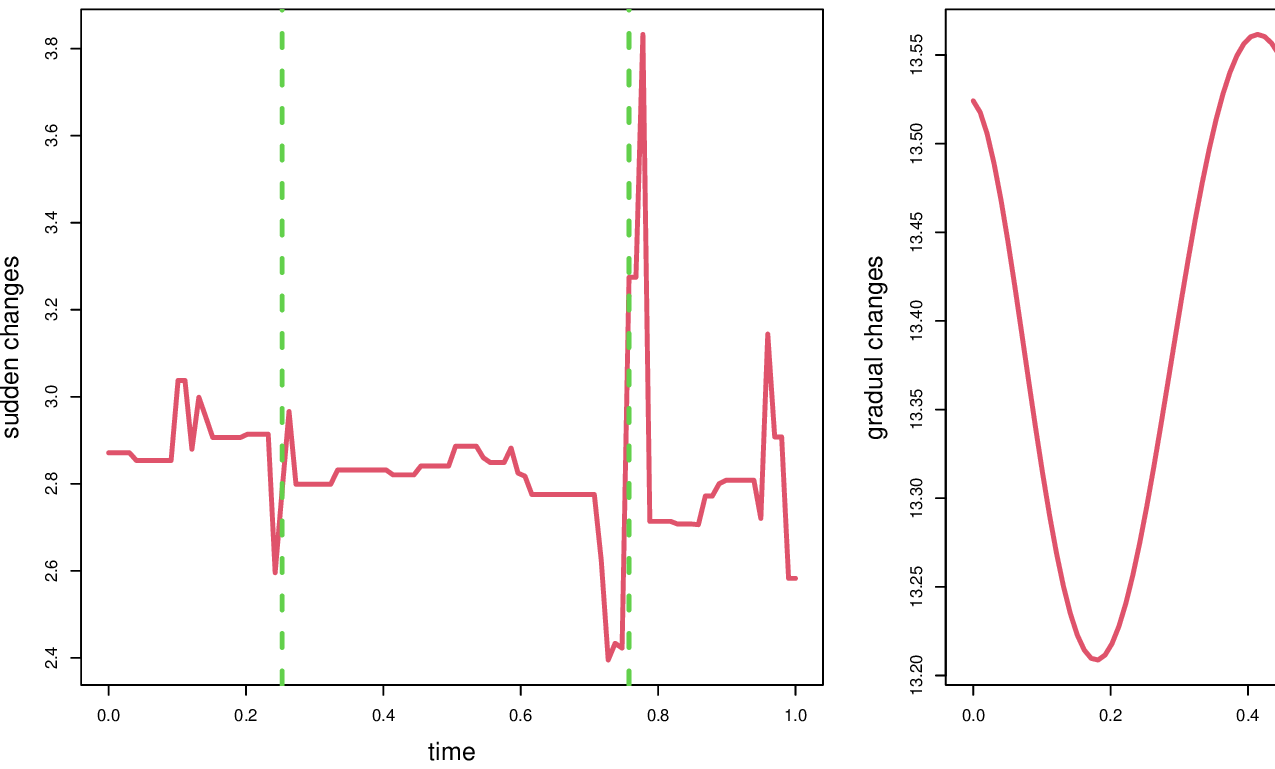} 
\caption{Sudden changes (left), gradual changes (middle), and input-output relationship (right). The true lines are shown as green dashed lines.}
\label{fig:plot_test}
\end{center}
\end{figure}

The left panel of Figure \ref{fig:plot_test} shows the estimate of sudden changes ($\bm U\hat{\bm e}$) over time. The tune-up locations are shown in this plot as dashed green lines. We can see that the data indeed show some sudden changes at these locations. Thus, further investigation of the reasons behind sudden changes at these time points may reveal the effect of tune-ups to the investigator, which can be an important discovery. The middle panel of Figure \ref{fig:plot_test} shows the estimate of gradual changes. Unfortunately, this does not provide much insights into the degradation effect on the laser energy. However, in this particular problem, the discovery of tune-ups, might lead the investigator to think about the possibility of degradation of the laser energy. The prediction of yield with respect to the laser energy is plotted as a red line in the right panel of Figure \ref{fig:plot_test}, which closely matches with the true relationship shown as a dashed green line.

The results are encouraging for the methodology to identify sudden changes in the data. Although only a single design parameter, the laser energy, was varied in this study, it is an important first step in understanding if the method may be applied to better understand sources of variation in MagLIF performance. In particular, the relatively low shot rate at Z ($\lesssim 1$ per day), the fact that the facility is shared across multiple experimental platforms, as well as the complex and evolving high-dimensional design space of MagLIF mean that repeat experiments are quite rare. It is left to future work to understand the limitations of the presented methodology in regards to requirements on experiment replication and design space explored.

\section{Conclusions}
Using the time of data collection as a surrogate variable, we have proposed a Bayesian framework to discover unknown variables that impacts the output. Our procedure only points to two types of unknown variables -- one causing gradual changes and the other causing sudden changes. Follow-up investigations are necessary to pinpoint the real variables that cause these changes. Our hope is that by classifying the numerous unknown variables into two groups and providing the time at which sudden changes occurred in the system are beneficial to the investigator to identify the real causes, which otherwise would be an extremely difficult task.

The proposed statistical model is complex with too many parameters and may have identifiability issues. As we have demonstrated through simulations, having replications and adhering to proper randomization rules will mitigate the identifiability issues and help in estimating the parameters accurately. Nevertheless, we admit that discovering the unknowns is a difficult problem and the work presented here should be viewed only as a first step in the discovery process.

Several improvements to our proposed methodology could be attempted in the future. We have chosen to ignore the uncertainties in the estimates of $\bm e$ in order to devise a computationally tractable algorithm. Incorporating these uncertainties would be a useful extension of the proposed method. Furthermore, we have used an additive structure for the sudden changes in our proposed model. It is possible that they interact with the other variables, which was the case in the MagLIF simulations. However, it is not clear to us how to extend our model to incorporate those interactions and still develop a computationally tractable algorithm for estimation. This could be another topic for future research.

\vspace{.2in}
\noindent{\Large\bf Acknowledgments}

\noindent This work is supported by an LDRD grant from Sandia National Laboratories. Sandia National Laboratories is a multimission laboratory managed and operated by National Technology \& Engineering Solutions of Sandia, LLC, a wholly owned subsidiary of Honeywell International Inc., for the U.S. Department of Energy’s National Nuclear Security Administration under contract DE-NA0003525. This paper describes objective technical results and analysis. Any subjective views or opinions that might be expressed in the paper do not necessarily represent the views of the U.S. Department of Energy or the United States Government. SAND2023-10721O.

\bibliography{bibliography}

\end{document}